\documentclass{article}

\usepackage{arxiv}

\usepackage[utf8]{inputenc}
\usepackage[T1]{fontenc}
\usepackage{microtype}
\usepackage{hyperref}
\usepackage{url}
\usepackage{booktabs}
\usepackage{multirow}
\usepackage{graphicx}
\usepackage{xcolor}
\usepackage{colortbl}
\usepackage{enumitem}
\usepackage[most]{tcolorbox}
\usepackage{natbib}

\setcitestyle{authoryear,round,citesep={;},aysep={,},yysep={;}}

\definecolor{darkblue}{rgb}{0, 0, 0.5}
\hypersetup{
  colorlinks=true, citecolor=darkblue, linkcolor=darkblue, urlcolor=darkblue,
  pdftitle={The Agentic Web Requires New Normative Infrastructure},
  pdfauthor={Cameron Pattison, Matthew Boulos, Noam Kolt, Changbai Li, Tiziano Piccardi, Seth Lazar}
}

\renewcommand{\headeright}{Preprint}
\renewcommand{\undertitle}{Preprint}
\renewcommand{\shorttitle}{Agentic Web}

\title{The Agentic Web Requires New Normative Infrastructure}

\author{
  Cameron Pattison \\
  Department of Philosophy \\
  Vanderbilt University \\
  \texttt{cameron.pattison@vanderbilt.edu} \\
  \And
  Matthew Boulos \\
  Imbue AI \\
  San Francisco \\
  \texttt{matt@imbue.com} \\
  \And
  Noam Kolt\\
  Faculty of Law and School of Computer Science and Engineering \\
  Hebrew University \\
  \texttt{noam.kolt@mail.huji.ac.il} \\
  \And  
  Changbai Li \\
  School of Electrical Engineering and Computer Science \\
  Oregon State University \\
  \texttt{lc@oregonstate.edu} \\
  \And
  Tiziano Piccardi \\
  Department of Computer Science \\
  Johns Hopkins University \\
  \texttt{piccardi@jhu.edu} \\
  \And
  Seth Lazar \\
  School of Government and Policy \\
  Johns Hopkins University \\
  \texttt{slazar@jhu.edu}
}

\date{\today}

\begin{document}

\maketitle

\begin{abstract}
The agentic web, in which users interact with the internet largely through agents acting on their behalf, is now technically feasible. However, many of the consumer and social benefits that could be realized by online AI agents acting scrupulously in their principals' interest are currently obstructed by outdated laws, terms of service, and other less formal practices which allow online platforms to block and degrade agent access, often in secret. Few distinctions are currently drawn between ``malicious bots'' and AI agents acting with the express delegated authority of a user. For the agentic web to realize its promise, it needs not only the technical infrastructure of protocols and interfaces, but the \textit{normative} infrastructure of a broadly-accepted and socially-beneficial set of laws, norms and practices governing agentic access to online properties. Building that normative infrastructure requires a society-wide conversation. This paper aims to help precipitate that conversation, to identify normative principles that can guide it, and to advocate for policies that enable users' appropriately delegated agents to act online on their behalf, with as few curbs on their doing so as is reasonable given the other legitimate interests at stake.
\end{abstract}

\section{Introduction}

We are in the middle of a profound shift in the digital economy. Over the last two decades the once-open internet evolved into an archipelago of walled gardens, each governed by a platform that offers some (often modest) measure of security and functionality in exchange for a growing degree of surveillance, control, and value-extraction \citep{Gorwa12052019, lazar_governing_2025,whitt_reweaving_2024}. AI agents powered by Large Language Models have the potential to disintermediate those platforms --- to hop over their garden walls, and in doing so to significantly enhance their users' autonomy \citep{marro_llm_2025,kapoor_build_2025, south_authenticated_2025}. 

Until recently, AI agents' execution of web-based tasks was constrained by technical capability \citep{kapoor_holistic_2025}. They could search the web for information \citep{wei_browsecomp_2025}, and support communication and coordination with a human in the loop, but they fell far enough short of being able to do all the things users might want them to, that the question of whether they \textit{should} perform those tasks was moot \citep{rabanser_towards_2026,kapoor_holistic_2025,xue_illusion_2025}. Since late 2025, this has changed \citep{OSWorld,time-horizon-1-1,openclaw_ai_browser_2026,anthropic_opus_4-5_2025,anthropic_opus_4-7_2026}. Significant advances in underlying model capability \citep{anthropic_opus_4-5_systemcard_2025} mean that agents are already undertaking tasks that were previously infeasible \citep{casper_ai_2025,staufer_2025_2026,rabanser_steverabhal-harness_2026, kapoor_holistic_2025}, and we plausibly have a clear line of sight to solving the remainder \citep{time-horizon-1-1}. AI agents should soon be able to work as delegated authorities that search, negotiate, transact, and otherwise coordinate on their principals' behalf across much of their online lives. The fundamental bottlenecks to the agentic web are no longer technical. They are instead normative: agents' functionality will largely be determined by whether or not \textit{other} digital intermediaries allow them access to content and actions on a user's behalf. It is therefore urgent for societies to decide: to what extent should users be entitled to delegate online actions to AI agents?

At present, we lack an intentional answer to this question. Existing terms of service, laws, and practices that long antedate AI agents permit online platforms and cloud security providers practically unlimited discretion to block agent access. Agents are often treated, indiscriminately, as ``malicious bots.'' Developers, meanwhile, are using all their technical expertise to evade these restrictions, resulting in an arms race in which the two sides' fortunes ebb and flow, and a non-trivial amount of societal value is burned up in the process. This ``cold war'' between platforms and AI agent developers has turned hot in a few notable cases, where platforms have sued agent developers to prevent them from accessing their services \citep{united_states_district_court_for_the_northern_district_of_california_amazoncom_2025,grynbaum_times_2023,longpre_consent_2024}. The agentic web requires new normative infrastructure. In this paper, we make three contributions to designing it.

 \begin{figure*}[t]
    \centering
    \includegraphics[width=\linewidth]{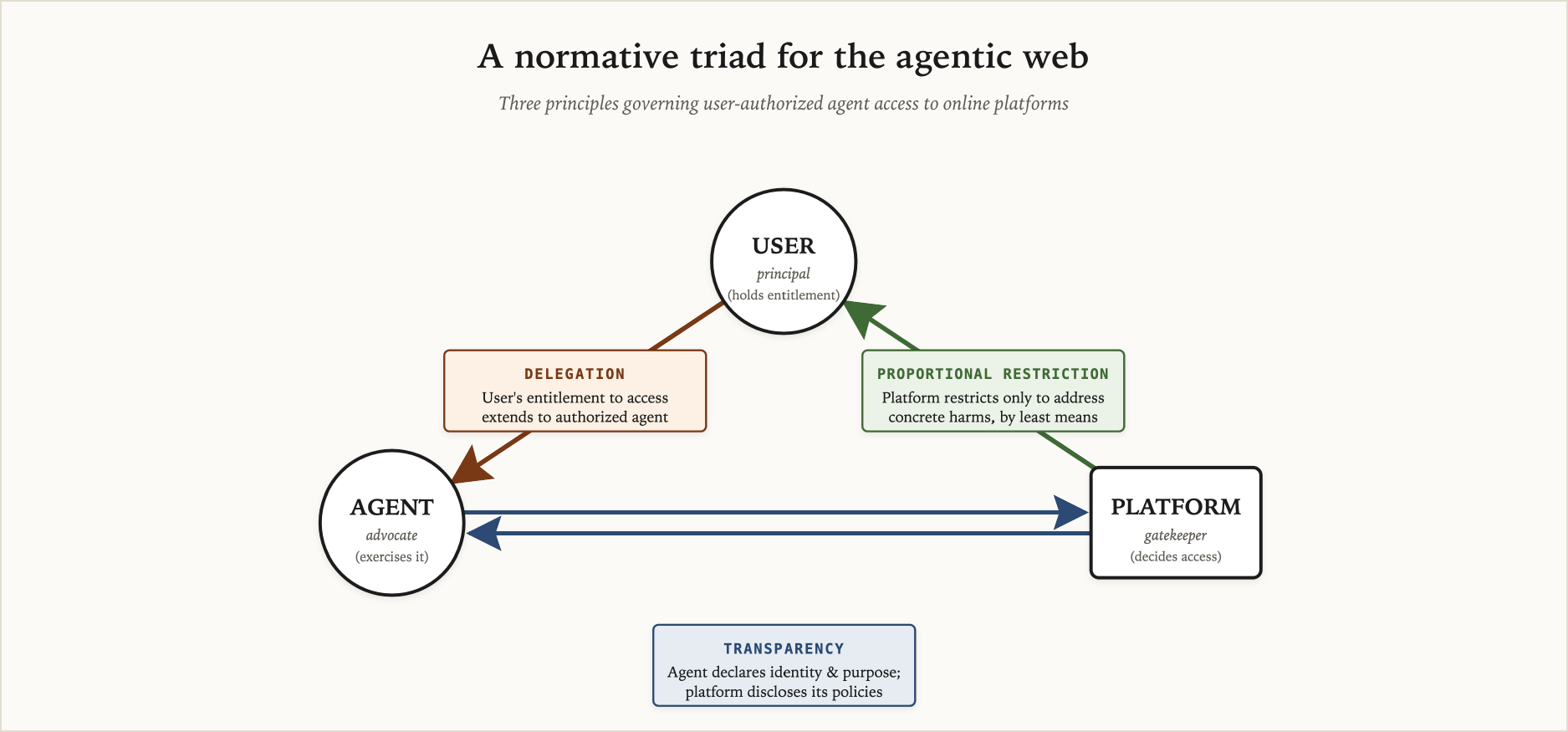}
    \caption{A normative triad for the agentic web. The user delegates authority
      to an agent (\textsc{Delegation}); the agent declares its identity and
      purpose to the platform, which in turn discloses its access policies
      (\textsc{Transparency}); and the platform restricts access only to address
      concrete harms by the least restrictive means
      (\textsc{Proportional Restriction}). These principles structure a regime
      that distinguishes user-authorized agents from indiscriminate bots.}
    \label{fig:triangle}
\end{figure*}

\textbf{First}, we argue that instead of this subterranean struggle between platforms and agent developers, we must urgently hold an open societal conversation about the nature and degree of access that AI agents should have to online properties. The future of the agentic web should be debated publicly before it is locked in privately. \textbf{Second}, we defend three principles that can help to anchor the conversation, from the perspective of the consumers who stand to benefit from AI agents. These are: \emph{delegation} (users entitled to access a service should ordinarily be able to exercise that access through user-authorized agents), \emph{transparency} (platforms should disclose how they treat agents, and agents should identify their purpose and authorizing principals), and \emph{proportional restriction} (platform restrictions should address concrete harms by the least restrictive means available). \textbf{Third}, we identify reasonable light-touch regulatory steps that could be taken, in light of those principles, towards normative infrastructure for the agentic web that better serves consumers' interests and autonomy.\footnote{We focus on the U.S. context because it is home to many of the major platforms and because its regulatory landscape is most influential for these platforms.} We propose these not as our final word on the societal debate that needs to occur about AI agent access, but as a demonstration that an alternative regulatory settlement that better serves consumers is feasible.

In what follows, \S\ref{sec:SoA} describes the technical and regulatory landscape of early conflicts over agent access. \S\ref{sec:new} explains why AI agents are meaningfully different from earlier access technologies. \S\ref{sec:Normative_principles} sets out three principles that should inform the design of normative infrastructure for the agentic web, \S\ref{sec:a_feasible_regulatory_path} sketches a feasible regulatory path, and \S\ref{sec:alternative_views} addresses the strongest objections.

\section{The Current State of Affairs}
\label{sec:SoA}
A technical arms race between platforms and AI agent providers has created an unstable de facto settlement that governs AI agent access to the web (\S\ref{sec:arms_race}). This settlement is substantially grounded in outdated legal precedents designed to regulate traditional web scraping (\S\ref{sec:regulatory}). 

\subsection{The Arms Race}
\label{sec:arms_race}

Automated access to the web pre-2024 was effectively regulated by two institutions: voluntary adherence to the Robots Exclusion Protocol \citep{koster_robots_2022} implemented through \texttt{robots.txt}; and a stack of protections aimed at identifying signals that distinguish automated clients from human users. Fingerprinting automated clients using user-agent strings, IP-address reputation, headless-browser traces, and inhuman request rates was effective and most scrapers respected --- and still do respect --- \texttt{robots.txt} preferences. This first wave of defenses was quickly overcome by early AI agents, such as PerplexityBot, which disregarded \texttt{robots.txt} and obscured the fingerprints that enable client blocking \citep{corral_perplexity_2025,knight_perplexity_2024,kim_scrapers_2025,bort_how_2025}. This led to widespread reports of massively increased traffic \citep{bort_how_2025,kim_scrapers_2025,belanger_ai_2025, mueller_how_2025,miller_new_2025,cloudflare_googlebot_2025,datadome_2025_2025}, and concerns about intellectual property theft by scrapers searching for LLM training-data \citep{grynbaum_times_2023,kim_scrapers_2025,liu_somesite_2024,longpre_consent_2024}. The predictable consequence was a loud denouncement of \emph{all} automated traffic as malicious \citep{bocharov_declare_2024} and further attempts to shutter AI agent access \citep{longpre_ai_2025,roose_data_2024}. 

But rapidly advancing AI agents present a moving target.  The move from CURL and Wget to competent browser use by AI agents undermined website providers' ability to distinguish between AI agents and users. AI browser control initially received a mixed reception and relatively light use \citep{openai_introducing_2025,manus_manusai_introducing_2025,anthropic_introducing_2024}, but today's equivalents such as \citet{openclaw_ai_browser_2026} offer users fast and efficient ways to bypass controls and reach much of the web content available to the users themselves. This includes paywalled and other authenticated content, which is accessed through direct agent control over already authenticated Selenium or Chromium-based browsers \citep{openclaw_ai_browser_2026}, by using authenticated cookies provided by users \citep{calzavara_measuring_2021,montulli_http_2000,malwarebytes_cookie_2026}, or by using plugins on Chrome browsers \citep{smooth_brain_browser_2026,randynamic_browserfly_2026}. These and other techniques --- BrowserBase integrations, CAPTCHA solvers, etc. --- have dovetailed with advances to agentic coding software like Claude Code and Codex to lower the technical barriers to users who want to release their AI agents onto an increasingly restricted web and bad actors impersonating those users through residential proxies and the like. Additional human gates like Cloudflare's agent-blocking Turnstile --- free to \emph{everyone} \citep{cloudflare_cloudflare_2023} --- have grown in popularity, but as soon as a defense is erected, means of circumvention soon follow.   

As technical barriers become less effective, those looking to defend their websites from automated traffic have turned to legal action and non-technical deterrence. \citet{nikita_bier_nikitabier_we_2026}, head of product services at X, threatened unilateral bans on any users experimenting with OpenClaw style automated access. Amazon, WhatsApp, and the New York Times are just a few of the major platforms to introduce terms of service clauses that restrict automated access \citep{whatsapp_whatsapp_2026,new_york_times_terms_nodate,amazon_conditions_2025}; many more are following suit \citep{roose_data_2024,longpre_ai_2025,longpre_consent_2024}. On the legal front, a suite of cases originally designed to prevent training-data scrapers  \citep{brittain_court_2026,united_states_district_court_for_the_northern_district_of_california_amazoncom_2025,grynbaum_times_2023} are still pending remedies and the groundwork is laid for future lawsuits under the statutes discussed next.

\subsection{The Regulatory Landscape}
\label{sec:regulatory}

Legal frameworks developed to regulate traditional scrapers now stand ready to govern AI agent access to the web. We focus on three frameworks that reach broadly across automated traffic, AI agents included: the Computer Fraud and Abuse Act (CFAA), website terms of service, and the common law doctrine of trespass. Other legal regimes, such as intellectual property law, are extremely important to scraping conversations, but have less purchase on AI agent access.\footnote{U.S. State and European privacy and security regimes (GDPR, CCPA/CPRA, DPA commissions etc.) exert significant pressure to restrict scraping in order to protect consumer privacy \citep{noauthor_art_nodate}. Additional complications in the European context also restrict blanket scraping under special circumstances \citep[see especially][]{noauthor_art_nodate,noauthor_digital_2026,noauthor_data_nodate}.}

\textbf{CFAA and anti-circumvention.} Barring a breach of basic rights protected by other statutes, platforms have wide latitude to impose technical controls. Once such controls are established, platforms can invoke the Computer Fraud and Abuse Act in the event of their circumvention \citep{united_states_department_of_justice_computer_2015}. The scope for ``exceeding authorized access'' was narrowed in \emph{Van Buren v. United States} \citep{supreme_court_of_the_united_states_van_2021}, where the court ruled that excession amounted to accessing areas of a computer one is not entitled to access. The Ninth Circuit’s decision on remand in \emph{hiQ Labs v. LinkedIn} bounded CFAA applicability further by clarifying that logged-out access could not constitute access ``without authorization'' \citep{united_states_district_court_for_the_northern_district_of_california_hiq_2022}. Nevertheless, this statute is highly relevant to agent access and is undergoing review now as seen in the preliminary injunction in \emph{Amazon.com Services LLC v. Perplexity AI, Inc.} \citep{united_states_district_court_for_the_northern_district_of_california_amazoncom_2025} where a conjunctive standard for ``authorization'' requiring both user authorization and platform authorization is forming.

\textbf{Terms of service as private regulation.} Contractual terms of service enable platforms to prohibit AI agent access by including clauses like ``no bots,'' ``no automated access or scraping,'' ``no credential sharing,'' and ``no use of content for machine learning or AI'' in terms of service (see \citep{amazon_conditions_2025,linkedin_prohibited_nodate,ebay_user_nodate,new_york_times_terms_nodate} for example). These clauses operate \emph{alongside} technical controls, enabling platforms to sue those who evade their technical controls \emph{and} breach their terms of service. \emph{HiQ Labs v. LinkedIn} \citep{united_states_district_court_for_the_northern_district_of_california_hiq_2022} provides the legal backing insofar as this case, while allowing the public-data scraping, affirmed the right of platforms to block automated access if such access was banned in the terms of service. \emph{Meta Platforms v. Bright Data} \citep{united_states_district_court_for_the_northern_district_of_california_meta_2024} affirms the boundary between public and private data by allowing ``logged-off'' scraping.

\textbf{Trespass and related residual doctrines.} In rare cases, common-law doctrines such as trespass to chattels may be invoked. This doctrine, designed to protect property owners' right to their own property from third party deprivation or prevention, is relevant when the platform can show damaging impacts of scraper activity requiring ``cost of repair or loss of rental'' \citep{wex_definitions_team_trespass_2023}. In \emph{eBay v. Bidder's Edge} \citep{united_states_district_court_for_the_northern_district_of_california_ebay_2000}, the court enjoined an auction aggregator whose crawler generated about 1.5\% of eBay’s daily traffic, treating it as trespass to chattels. However, \emph{Intel v. Hamidi} limits the application of trespass insofar as there, the court denied Intel's claim that Hamidi trespassed against it on grounds that the defendant neither damaged nor impaired Intel's servers in sending mass emails tens of thousands of Intel employees \citep{california_supreme_court_intel_2001}. Stories about DDoS-like AI agent traffic \citep{bort_how_2025,kim_scrapers_2025} alongside industry reports about how automated access is damaging platforms \citep{mueller_how_2025,miller_new_2025} strengthen the \emph{potential} of trespass claims, but litigation under this doctrine remains rare and difficult given the burden of proof placed on platforms to demonstrate harm or the reasonable expectation thereof.

\section{Why AI Agents Are Meaningfully New}
\label{sec:new}

The technical and regulatory measures surveyed in \S\ref{sec:SoA} were designed for traditional crawlers and scrapers. Extending them to AI agents is apt only if AI agent browsing is relevantly similar to those earlier technologies. We argue that these are indeed very different technologies. Most importantly, where earlier bots were largely a net negative for society (including for consumers), properly delegated AI agents could be enormously individually and socially beneficial.   

Web scrapers and other bots are (relatively) dumb, deterministic programs that perform a single, simple function. When you ask an agent to help you choose a restaurant for the night, it may scrape reviews from many web-pages, or delegate that scraping task to a sub-agent. So it may sometimes be true that AI agents \textit{use} bots. But the agent itself is something much more sophisticated.  At their best, AI agents can act, negotiate, transact, and coordinate on users' behalf \citep{shavit_practices_nodate, lazar_frontier_nodate, gabriel_ethics_2024, marro_permission_2026,krier_coasean_2025,anthropic_project_2026,NBERc15309}. They can maintain persistent context, form relationships with services, and adapt strategies over time \citep{marro_permission_2026,kapoor_build_2025}. They can exercise judgment --- choosing among options, timing actions, weighing tradeoffs --- in ways that scripts and scrapers cannot \citep{anthropic_project_2026,krier_coasean_2025,NBERc15309,lazar_moral_2025}. And they can evade automated measures intended to constrain them, even avoiding patterns of mouse and scrolling behavior that would give away earlier generations of bots. 

Laws, norms and practices designed for one technology generally should not be unthinkingly extended to a new technology that, though superficially similar, is in fact quite different. These technical differences tend to underpin normative differences. First, earlier bots, for the most part, would not use the internet on behalf of and in much the same way as the person or company that deployed them. They would do something that no human user could do on their own. Today's AI agents of course \textit{can} operate in this superhuman way. But they can also be narrowly constrained to performing only tasks that their principal would be able and entitled to perform. Second, while some earlier bots were broadly socially beneficial --- for example indexers, archivists, and performance monitors --- most other bots (especially those that might ignore \texttt{robots.txt}) were very clearly not. They were likely to involve, for example, competitor price-scraping, data harvesting, content scraping for SEO optimization, and automated nuisance messaging \citep{imperva_2024_2024}. They typically represented avenues for some fairly unscrupulous company to take advantage of poor enforcement for the sake of  a private profit.  

Today's AI agents promise much more individual and social benefit than that. Most obviously, capable agents can save users substantial amounts of time by searching, comparing, summarizing, and acting across the web on their behalf \citep{gabriel_ethics_2024, time-horizon-1-1,kwa_measuring_2026}. More ambitiously, they could make forms of assistance now available mainly to the wealthy --- lawyers, consultants, brokers, and personal assistants --- available to ordinary users across a wide range of online tasks \citep{anthropic_project_2026,krier_coasean_2025}. A user-authorized agent could compare prices, negotiate terms, monitor opportunities, challenge unfair charges, manage subscriptions, or coordinate complex administrative tasks. In doing so, agents could shift the balance of power between individual users and the platforms, firms, and institutions that currently set the terms of most online interaction \citep{kapoor_build_2025}.

Because agents can operate across services, they can make interoperability practical from the user's side rather than waiting for platforms to provide it \citep{marro_llm_2025}. They can lower switching costs, reduce dependence on platform defaults, and help users navigate around walled gardens \citep{kapoor_build_2025, lazar_moral_2025, marro_llm_2025}. This also matters for attention. Much of the contemporary web is designed to capture, steer, and monetize user attention; social media recommendation systems are only the most visible example \citep{kang_meta_2026,kang_meta_2026b,manuali_addictive_2025}. Properly aligned agents potentially offer a different interface: one organized around users' reflectively endorsed goals rather than platform-optimized engagement \citep{lazar_lecture_2024,schuster_attention_2025}. That promise is not without risks, especially where agents mediate news, recommendations, or politically salient information \citep{gabriel_ethics_2024}. But the relevant comparison is between different systems for allocating attention, only some of which are accountable to the user --- not between agent mediation and a neutral status quo.

User-authorized agents could expand users' collective power. Many online harms persist because individuals lack the time, information, or coordination capacity to respond effectively. Agents acting for users could help overcome these constraints by identifying common grievances, coordinating exit or complaint, comparing boilerplate terms, or helping users resist forms of ``digital resignation'' \citep{draper_corporate_2019,radin_boilerplate_2013}. Of course, these goods are unlikely to come unalloyed. Multi-agent systems may create risks of collusion, cascading failure, or emergent behavior, and aligning agents reliably with users' interests remains a central technical and institutional challenge \citep{kapoor_build_2025,motwani_secret_2024,hammond_multi-agent_2025}. But with appropriate governance, agent advocates could give users new leverage in digital environments where individual choice is often formally available but practically weak \citep{kapoor_build_2025,radin_boilerplate_2013}. Blocking agent access constitutes a significant setback to consumers' interests. 

Lastly: it is important to recognize the potential scope of user delegation to AI agents. It reaches far beyond just the allocation of online attention, or the consumption of content. Many people have found tremendous value in using agents as intermediaries to the other digital services that they use, as well as the various forms of context that they generate across Software as a Service properties. The individual and social benefits of AI agents extend to every domain in which having a smart, autonomous delegate acting on your behalf is empowering.

This leads to our first conclusion. The existing laws, norms and practices that govern automated access to online properties were designed for an intrinsically different technology than today's AI agents, with a different connection to users' antecedent sphere of permissible action, and different social costs and benefits. Minimally, these changes demand a reopened societal conversation about which automated services are allowed to access which online properties.

\section{Normative Principles}
\label{sec:Normative_principles}

We propose three principles for governing agent access to platforms: \textit{delegation} (\S\ref{sec:delegation}), \textit{transparency} (\S\ref{sec:transparency}), and \textit{proportional restriction} (\S\ref{sec:proportional_restrictions}). We are not trying to identify and balance all the relevant normative considerations. Instead, we are identifying three interrelated principles that should be part of any such account. 

\subsection{Delegation}
\label{sec:delegation}
\textbf{If a user is entitled to access content or a service, they should be permitted to delegate that access to an appropriately authenticated agent acting on their behalf, and there should accordingly be a strong presumption against that access being blocked.}

This principle rests on an established feature of delegated authority: when a principal exercises an entitlement through an authorized delegate, the delegate does not assert or require a new entitlement of its own. It acts as the principal’s means of exercising an entitlement the principal already holds \citep{hobbes_leviathan_1988,locke_second_1980}. Delegation theory has long recognized this structure. In bioethics, for example, surrogate decision-makers preserve a patient’s decisional authority in circumstances where direct exercise is impossible or impracticable: the surrogate carries the patient’s authority into a context the patient cannot navigate alone \citep{buchanan_deciding_2004}. To block the surrogate, on that view, is to block the patient’s own exercise of authority.

The digital case is structurally similar: a user principal is delegating her powers to an AI agent, thus transferring her authority and exercising her autonomy. Now, one could counter at this point that platforms are entitled, at least in principle, to prevent users from delegating their access to some online property to a \textit{human} third party, so why shouldn't they be able to do the same for agents? However, the obvious reason to prevent delegation of access authority to other humans is that the human agent may thereby derive a benefit that they would otherwise have had to pay for. AI agents, by contrast, do not derive private benefit from accessing content on their users' behalf, and would not, were they not acting for the user, be potential customers whose business is being lost by permitting delegation. 

What is more, if a user is entitled to access a platform, but wants to use an agent to do so, in order to pursue some other activity themselves, then the agent is effectively the tool --- in this case, the browser --- chosen by the user.  Just as platforms should not make your access to their content conditional on the specific browser that you use to do so, the same basic neutrality principle should allow you to use an AI agent instead \citep{wu_network_2003}. The entitlement remains the user’s; only the mode of exercise changes. So long as the agent acts within a user-authorized scope, for user-defined purposes, and subject to limits that the user would also be constrained by, its activity should not be treated as a separate access claim by an independent third party. 

Agency law reinforces this structure. As one of the oldest bodies of common law, it provides a mature framework for delegated action, including duties of loyalty, care, and obedience owed by agents to principals \citep{ali_restatement_agency_2d_v2_1958,kolt_governing_2024,benthall_designing_2023}. Importantly, agency law also puts boundaries on delegated action, marking where a delegate acting on behalf of a principal oversteps their authorization and so violates the principal-agent relationship \citep{ali_restatement_agency_2d_v2_1958}. This standard is important for AI agents insofar as it conditions the agent such that an action that stands beyond delegated scope --- to scrape training-data for its model provider, for instance --- violates the principal-agent relationship and constitutes wrongdoing. Authenticated delegation protocols extending OAuth 2.0 and OpenID Connect stand as open paths for operationalizing formal, auditable links between principal and agent \citep{south_authenticated_2025}.

\subsection{Transparency}
\label{sec:transparency}

\textbf{Platforms should be required to disclose how they handle agents --- what access they permit, what they block, and on what grounds. Agents, in turn, should be required to identify themselves and their authorizing principals to the platforms they access.}

Transparency is a precondition for the autonomous choice the delegation principle protects: a user cannot meaningfully decide which agent to use, or whether to remain on a platform that frustrates her chosen agent, when the relevant facts are concealed. The disclosure obligations from fiduciary, consumer-protection, and communications-regulation traditions all address the same conditions present here including but not limited to asymmetric information, costly switching, and post-sale conduct invisible to the buyer. Transparency also enables market discipline: if users know which platforms block their agents, they can switch \citep{kapoor_build_2025}. Covert blocking by platforms means that users cannot tell their agent is being degraded. Transparency requirements are politically feasible, precedented (FTC disclosure mandates), and facilitate the operation of markets that advance consumer well-being.

Transparency is also bidirectional. Agent developers must ensure their agents self-identify \citep{romm_build_2025,south_authenticated_2025,chan_ids_2024,south_authenticated_2025}. While some go further, insisting that agents must also always identify their purpose and act within a single purpose \citep{romm_build_2025}, we subsume this standard under the act within delegated scope, outlined above. Without identification, platforms cannot distinguish a legitimately delegated agent from an unauthorized scraper: an agent that accesses paywalled content with valid user credentials but no identifying metadata is indistinguishable from a malicious bot. Agent identification is a precondition for the delegation principle to function, since platforms cannot be expected to honor delegated access when they cannot verify the identity of the agent and of the agent's principal. Again, this is technically feasible through an extension of existing technical components, including OAuth 2.0, OpenID Connect, W3C Verifiable Credentials, and Decentralized Identifiers \citep{south_authenticated_2025, chan_ids_2024}. 

An interesting question arises when a user would be entitled to anonymously access some online content. According to the delegation principle, if the user can permissibly perform some action, the user's agent should presumptively be able to do so too; but the transparency principle enjoins the agent to identify itself and on whose behalf it is acting. We suggest that for these cases, agents should be able to identify themselves \textit{as authorized}, without explicitly identifying the actual individual by whom they are authorized by default.

\subsection{Proportional Restriction}
\label{sec:proportional_restrictions}

\textbf{Platform restrictions on agent access must be proportional to concrete risks.}

This principle reflects a final, familiar constraint: a private actor that exercises power over others' lawful conduct may impose restrictions only when needed and in the least restrictive form that addresses the underlying concern \citep{pettit_republicanism_2003,kolodny_pecking_2023,kolodny_rule_2014}. By \emph{concrete risks} we mean probabilities of harm grounded in evidence about the agent class in question --- whether \emph{empirical} (incidents, audits, observed misuse) or \emph{anticipatory} (red-team and safety-evaluation findings that establish a credible mechanism of harm). We adopt a common standard of legal proportionality \citep{barak_proportionality_2012}: a restriction must be (a) rationally connected to the stated aim --- it must be well suited to actually advancing that aim, (b) necessary --- no less restrictive alternative achieves the same aim, and (c) proportionate stricto sensu, inasmuch as its costs do not outweigh its benefits.

Legitimate grounds for restriction may include server load and infrastructure costs, fraud and abuse prevention, safety (e.g., agents purchasing controlled substances), privacy protection of third parties, and \emph{prevention of agent conduct that would be unlawful if conducted by the principal} --- such as insider trading, fraudulent misrepresentation, unauthorized practice of regulated professions, copyright infringement, or computer-intrusion offenses --- where the restriction targets the specific unlawful conduct rather than the class of user-authorized agents at large \citep{kolt_legal_2026, kolt_superintelligence_2026}. We understand also that there may be spaces on the internet --- support groups and other forums for humans for instance --- that are in core tension with an influx of AI agents. We count the preservation of these spaces among the legitimate grounds for restriction, though the invocation of a special space should be treated as exceptional requiring concrete justification. 

Plausibly illegitimate grounds include maintaining data lock-in, protecting competitive position, and speculative ``what if'' scenarios. Restrictions that are justified in the language of security, fraud prevention, or infrastructure should not be used as pretexts for preserving lock-in, self-preferencing a platform’s own agent, or foreclosing rival agent ecosystems \citep{kapoor_build_2025}. The point of proportionality is both to protect users from overbroad restrictions in individual cases, and to prevent platforms from converting control over access into a competitive moat.

The preservation of ad-revenue may also turn out to be illegitimate grounds for blocking. Platforms will likely insist that agent-mediated access circumvents ad-supported monetization, which itself supports the economic model of the free web. However, the law allows that users may dictate how they render web content --- in particular, that they may use ad-blocking extensions and web browsers \citep{miller_legal_2018}. User-authorized AI agents --- with respect to ad-revenue justifications --- would be functionally equivalent. Preserving a specific UI-based advertising model would be inadequate grounds for banning them. Rather than allowing platforms to unilaterally block the pathway to a better web, we instead must start to think about how the economics of the agentic web should look (see\S\ref{sec:alternative_views}).

\section{A Feasible Regulatory Path: Legislation and Targeted Enforcement}
\label{sec:a_feasible_regulatory_path}

While our aim in this paper is to provide resources for a broader conversation about appropriate AI agent access to the web, rather than foreclose it with a presumptive solution, it is useful to consider concrete, politically feasible regulatory interventions that could achieve a more user-centered approach to AI agent access. In this section, we describe a menu of FTC and legislative options. FTC enforcement might support transparency and proportional restriction (\S\ref{sec:ftc}) and legislatures, both state and federal, might target interoperability, data-portability, and platform power more generally in broadly appealing forms (\S\ref{sec:legislative}).

\subsection{The FTC Path}
\label{sec:ftc}

Today, FTC authority operates within a very narrow scope. The Commission cannot simply declare a general right of agent access by rule: recent appellate vacaturs of major FTC rules underscore the present fragility of broad administrative rulemaking, at least absent exceptionally careful statutory and procedural grounding (see \emph{National Automobile Dealers Association, et al. v. FTC} \citep{nada_ftc_2025} and \emph{Custom Communications, Inc. v. Federal Trade Commission} \citep{custom_communications_2025}). Nor is ordinary antitrust doctrine a promising substitute. In \emph{Verizon Communications v. Law Offices of Curtis V. Trinko}, the court affirmed a general rule of ``no duty to aid competitors'' \citep{trinko_2004}. \emph{Aspen Skiing Co. v. Aspen Highland Skiing Corporation} stands, after \emph{Trinko}, as the boundary of Section 2 liability and the mold into which any antitrust enforcement must conform \citep{aspen_skiing_1985}. A platform blocking third-party agents will likely characterize the conduct as a unilateral access decision protected by \emph{Trinko} precedent rather than \emph{Aspen v. Aspen} exclusionary conduct. Still, there are ways that narrow FTC enforcement and rulemaking may be capable now and in the future of scrutinizing how platforms restrict user-authorized agents.

Section 5 of the FTC Act makes it unlawful for companies to engage in deceptive acts and practices. Deception as elaborated in the FTC Policy Statement on Deception \citep[appendix][]{cliffdale_associates_1984}, requires a misleading act, omission, or representation made by a company to a reasonable consumer whose decision is materially affected. If platforms purport to give unfettered access to their properties while quietly block agentic access \citep{belanger_ai_2025,cloudflare_trapping_2025}, they may thereby make a misleading representation of access that is not reasonably avoidable and which ``materially'' affects consumers' decisions about where to spend their subscription dollars. For a practice to be ``material,'' the ``basic question is whether the act or practice is likely to affect the consumer's conduct or decision with regard to a
product or service'' \citep[appendix][]{cliffdale_associates_1984}. As agentic access becomes more important to consumers, blocking becomes increasingly material. Failure to disclose silent throttling, degradation, or refusal-to-deal with user-authorized agents while leaving consumers in the dark about these practices is increasingly likely to effect which platforms users prefer to deal with, and thereby to effect consumer decisions about which platforms to purchase subscriptions to. As such, silent degradation may constitute an omission that materially affects a reasonable consumer.

Section 5 also makes it unlawful for companies to engage in unfair acts or practices that result in substantial injury to consumers that is not reasonably avoided, and that is not outweighed by countervailing benefits to consumers or competition \citep[see \emph{FTC v. Sperry \& Hutchinson} and \emph{Orkin Exterminating Co. v. FTC} alongside the 1980 FTC Policy Statement on Unfairness][]{us_supreme_court_federal_1972,orkin_exterminating_1988}. This prohibition stands free of the misleading representations and false advertising that conditions claims of deception. When consumers find their agents blocked, they may claim substantial injury to their interests that are not outweighed by countervailing benefits and, especially when blocking is covert, cannot be reasonably avoided by consumers.\footnote{Note that when restrictions are justified by concerns not limited to security, fraud-prevention, privacy, and infrastructure, unfairness cannot and should not be claimed. These concerns justify restriction on our \emph{Proportional Restriction} principle as well.} 
 
Self-preferencing and exclusionary conduct may also gain in relevance as AI agents proliferate. If platforms redesign their interfaces so as to make them incompatible with the agentic systems they are competing with, while at the same time offering little to no advantages for their users, they may act anti-competitively in ways analogous to C.R. Bard's ``improvement'' of their biopsy gun in \textit{C.R. Bard Inc v. M3 Systems }\citep{us_court_of_appeals_for_the_federal_circuit_cr_1998} and Keurig's improvement of their coffee machine in \textit{In re: Keurig Green Mountain Single-Serve Coffee Antitrust Litigation} \citep{southern_district_of_new_york_re_2025}. Bard's ``innovations'' were deemed anti-competitive because their innovations accomplished nothing but the exclusion of competitive, third-party needles and Keurig's similar redesign of their machines to reject third-party k-cups faces similar allegations in ongoing litigation \citep{hovenkamp_antitrust_2023}. Microsoft's behavior in \emph{United States v. Microsoft Corp.} \citep{us_court_of_appeals_for_the_district_of_columbia_circuit_united_2001} in which Microsoft made exclusionary decisions to preference their own web-browser over competitors ``where the exclusion made no economic sense for Microsoft \emph{but for the exclusion of rivals}'' \citep[][14]{jacobson_competition_2023} provide the digital, platform analogue to these older cases \citep[see also the \emph{Statement of the Federal Trade Commission Regarding Google’s Search Practices In the Matter of Google Inc.}][]{federal_trade_commission_statement_2013}. Platforms that innovate \emph{only} to exclude third-party agents, \emph{without} rendering benefits to users, fail even the most platform-friendly, ``no economic sense [except to exclude competitors] test'' \citep{werden_identifying_2006,katz_does_2025,jacobson_competition_2023} and thus participate in liable, exclusionary conduct. 

\subsection{The Legislative Path}
\label{sec:legislative}

Legislation, whether passed at the state or federal level, has noteworthy advantages over administrative rulemaking or other executive action. For one, while bodies like the FTC are restricted by precedent, legislators are free to propose new rules that break from the past and are thus more responsive to genuinely new phenomena. We argued in \S\ref{sec:new} that delegated AI agents are meaningfully new. Legislation is thus well positioned to write rules to guide their development and deployment. 

Among the principles discussed above, transparency-related requirements are the most plausible target for legislation. The aim is to ensure that platforms and user-authorized agents can recognize one another through common credentialing and identification standards, and that any restrictions on agent access are disclosed clearly rather than applied opaquely or discriminatorily. Legislation of this kind would create the basic conditions for delegated agent use without foreclosing legitimate platform interests in security, fraud prevention, privacy, or system integrity. A federal statute provides the cleanest and most uniform framework, though state-level action may be more tractable in the near term.

State or national legislation might also target blanket-blocking of automated access as a part of broader efforts to prevent platforms from funneling users into self-preferencing, platform-based agents. We share concerns voiced in \citet{kapoor_build_2025} about the emergence of platform agents that act in the interests of platforms, rather than in user interests. We are further concerned that platforms may short-circuit the interoperability potential in AI agents by blocking all outside access to their platforms, thereby pushing users toward their own, in-house AI agents. Such action lies close to the self-preferencing, platform dominance, and anti-competitive action that the American Innovation and Choice Online Act (AICOA) sought to curtail \citep{cicilline_american_2022,klobuchar_american_2022}. It may also complement or be complemented by the Augmenting Compatibility and Competition by Enabling Service Switching Act of 2025 (ACCESS Act) and related legislation that aims at preserving user-data portability and interoperability across platforms \citep{warner_augmenting_2025}.\footnote{This work also sits very close to Sen. Mark Warner's Artificial Intelligence Access, Gatekeeper Exchange, and Nondiscriminatory Transfer Act (AI AGENT Act) 2026 released in discussion draft form on June 29, 2026.} 

Any action along these lines to divest platforms of concentrated power may appeal to both sides of a divided legislative body at both the federal and state level. On the left wing, there is deep-seated concern over the degree to which platforms wield economic power over consumers --- as evinced in the AICOA and ACCESS Act. Ensuring that platforms cannot block a promising interoperability tool from the outset may speak to these and other interests on the left. On the right, interoperability promised by AI agents charts a path forward that, if done well, stands to diffuse platform ideological power. Instead of being locked-in to a platform that provides univocal information, consumers might instead lean toward a more balanced media diet. Of course, dangers persist here and are discussed at length in \S\ref{sec:alternative_views}. 

To date, a recognition of the importance of AI agents has been slow to materialize in state and federal legislatures. At the federal level, agentic AI has only recently gained acknowledgment, appearing in the House Commerce, Justice, Science, and Related Agencies Appropriations Report in May 2026 \citep{rogers_commerce_2026} and in the National Defense Authorization Act signed December 2025 \citep{119th_congress_national_2025}. It is still absent at the state level to the best of our knowledge. As a result, CFAA and copyright law are filling the legislative void --- we look particularly to \emph{Amazon v. Perplexity AI} as perhaps the most relevant case bearing on this domain \citep{united_states_district_court_for_the_northern_district_of_california_amazoncom_2025,brittain_court_2026}. This state of affairs calls for serious, broadly appealing legislative efforts so that the future of the web is not determined by the adjudication of laws not designed with agents in mind, or settlements made between platforms and AI providers without user interests in mind.

\section{Objections and Residual Risks}
\label{sec:alternative_views}

We aim to start a conversation about the future of AI agent access on the web. We have charted what many will see as a permissive path forward for this access; alternate views abound and several listed below \emph{not} in order of importance deserve special mention. These we divide into two rough categories, economic and platform concerns (\S\ref{sec:economic}), and concerns about AI agents and dangerous emergent phenomena related to their web-access (\S\ref{sec:agent_based_concerns}).

\subsection{Economic and Platform Concerns}
\label{sec:economic}

\subsubsection{Open Agent Access Imposes Costs and May Undermine Web Economics.}
\label{subsec:economics}

Platforms are right that agent traffic can impose server costs, complicate abuse detection, and degrade service quality \citep{datadome_2025_2025,cloudflare_googlebot_2025}. We have heard from platforms that agents responding to single user queries may result in 20x web-traffic compared to human users, and formal reports from sites like Wikipedia have documented significant increases in automated traffic \citep{miller_new_2025,mueller_how_2025} --- though this load is, today, still largely accounted for by training-data scrapers which our framework would not protect from bans \citep{datadome_2025_2025,cloudflare_cloudflare_2025}. 

Agents also threaten ad-based revenue models that are common across the internet: if agents intermediate attention \emph{instead} of sending humans to platforms and access the web through \texttt{http} requests that serve \texttt{html} code instead of full browser clients, advertising as it exists may not even render to such agents, let alone actually reach the eyes it is intended for \citep{miller_new_2025,chapekis_google_2025,nitu_machine-readable_2025}. This issue has gained serious public attention \citep{cohen_how_2026,knight_ai_2026,allyn_online_2025}. However, this concern rings rather hollow from large scale internet platforms that have themselves been widely accused of re-presenting content from online publishers, denying the latter revenue accordingly. The broad lesson is that we need to devise an economic model for the open web that supports a flourishing creative economy and information ecosystem. Advertising placed alongside content worked for a while; since online platforms have come to dominate user attention, it arguably has not. The only difference agents make is to disintermediate the incumbent intermediaries. 

Though we cannot offer a comprehensive economic blueprint for the agentic web, markets are already responding to this shift in monetization models. Several publishers have licensed content to AI providers \citep{bruell_marketplaces_2026} as a way of deriving revenue from training-data scrapers, as well as from the growing market for retrieval-augmented generation fetches: when a user asks ChatGPT, Claude, Gemini, and others for the answer to a question, these chatbots often search the web for up-to-date answers, and those positioned to serve these answers --- especially news and culture publishers --- are increasingly interested in selling access through these licensing deals \citep[see][]{bruell_exclusive_2025,bruell_openai_2024,bruell_news_2026,bruell_marketplaces_2026,reuters_openai_2024,davalos_openai_2024}. These revenue futures appear to be some of the strongest and most viable paths forward for online publishers, including platforms, in an AI-mediated world. 

AI agents in fact introduce many more monetization options than were previously feasible, because they will in principle be able to negotiate for access to content and make micro-transactions that directly fund content creators. Many of the biggest obstacles to these kinds of approaches to funding online publishers concern the time and cognitive effort required to judge whether a given piece of content is worth a given fee. Users will be able to outsource those judgments to agents \citep{lazar_moral_2025}. In addition, monetization schemes such as agent compatible pay-per-crawl schemes, or neutral marketplaces for licensing providers \emph{en masse} --- such as those provided by TollBit, RedPine, RSL, and others --- may better support an open, fair marketplace. 

Residual concerns that agent intermediation will undermine ad-revenue and, by extension, independent publishing so rapidly that alternative monetization schemes will not have time enough to mature, are handled by the proportional restriction principle, which cuts in two ways. It prohibits anti-competitive suppression of bots on the part of platforms, but also permits cost-based restrictions when platform needs are at stake. 

\subsubsection{Agent Access Will Be a Gift to Training-Data Scrapers.}
Within AI bot traffic, training-data scraping still dwarfs traffic from delegated agents (sometimes called ``user action'') or search bots \citep{belson_2025_2025}. By some estimates, training-data scrapers dominated AI bot traffic, producing as much as 32 times the traffic that user delegated agents produced at peak in 2025 \citep{belson_2025_2025}.\footnote{This data comes from Cloudflare's 2025 Year in Review and 2025 Radar, and so is limited to the websites sitting behind Cloudflare's protections \citep{belson_2025_2025,cloudflare_cloudflare_2025}.} This raises a concern that any framework aimed at letting in delegated agents will, in practice, launder the much larger flood of training-data scrapers. 

This kind of concern is what motivates the measures discussed in \S\ref{sec:Normative_principles}. Transparency (\S\ref{sec:transparency}) requires that agents identify themselves and their principals. Scrapers that evade controls and fail to identify themselves and their principal properly may be blocked. Delegation (\S\ref{sec:delegation}) requires that an agent act within the scope given by the principal and doing double-duty by acting for the principal's interests while also collecting training-data at the same time is a clear violation of delegation-scope. Identification protocols make these restrictions enforceable \citep{south_authenticated_2025,marro_scalable_2024,marro_llm_2025} and platform content delivery networks (CDNs) like Cloudflare already recognize that not all bots are bad bots, proposing their own means for recognizing properly identified and scoped agents \citep{meunier_moving_2026}. We provide a more detailed, user-centered account of that same distinction between good and bad bots in hopes that the next few months and years do not foreclose user access to the benefits promised by good bots. 

\subsubsection{Web Agents Fuel the ``Attribution Crisis.''}

Recent literature suggests that AI agents may fuel an ``attribution crisis.'' Agents often answer user queries without accessing the internet, and often cite only a fraction of the web-pages they visit when they do go online \citep{strauss_attribution_2025}. This trend risks plagiarism at scale and may undermine the production incentives for high-quality online content. It also makes branding much harder for well known publishers and platforms when answers come from chatbots, not from interaction with branded surfaces, and when online properties \textit{are} cited, citations risk the misattribution of LLM-based errors to publishers \citep{bbc_audience_2025,european_broadcasting_union_news_2025}. Facilitating AI agent access to the web may exacerbate these existing issues. 

These issues are serious and, if left unmitigated, a continued attribution crisis may push the web toward a radical fragmentation of information, harming users' information interests by making it unclear where information is coming from. Branding is a secondary issue to these epistemic ramifications borne by users. But a move toward a more decentralized web does not necessarily entail grave harm to independent publishing (see \ref{subsec:economics}). 

So there are at least two responses to the present attribution crisis. In the first place, we should pressure the companies that present AI agents to the public to maintain attribution as a central piece of user interface. If attribution disappears, decades of work in media literacy and users' general ability to trace the provenance of their information will disappear with it. Arguments for the protection of attribution should focus on preserving this resource \emph{for users}, with brand stability and the preservation of independent publishing as important contributing factors to protection of these essential user interests. Secondly, the intermediation of AI systems in general is bringing on these issues with attribution and provenance tracking, whether we turn toward AI agents as intermediaries or remain with chatbot interfaces in the coming years. Attribution is a serious and separable issue that our framework is not primarily aimed at, but which our framework is not in core tension with. We do not claim to solve all problems with the future of the internet here, and this is an important issue that requires further research. 

\subsubsection{Platforms Own Their Own Infrastructure and Should Not Be Obligated to Provide Access.}

Amazon, X, and others begin by reminding audiences that platforms own their content and infrastructure (see Nikita Bier's comments \citet{nikita_bier_nikitabier_we_2026} and Amazon's lawsuit against Perplexity \citet{united_states_district_court_for_the_northern_district_of_california_amazoncom_2025}). Allowing others to use this content and infrastructure may impose an improper ``duty to aid rivals'' which is not generally recognized. In addition, this duty, as was mentioned above, has been positively denied in U.S. law by the Supreme Court's opinion in \emph{Verizon Communications v. Law Offices of Curtis V. Trinko}. There, in rough outline, Verizon under the 1996 Telecommunications Act was required to share their infrastructure with competitors, thus breaking up the legal monopoly they had formed on services prior to the Act. Trinko sued Verizon on the grounds that Verizon's alleged violation of the Telecommunications Act was \emph{also} anti-competitive behavior cognizable under Section 2 of the Sherman Act. Trinko's suit failed. Without any such Telecommunications Act mandating access to platform infrastructure, any attempt by AI agent providers to open platform infrastructure to delegated agents should flounder on \emph{Trinko} precedent. 

Now, our proposal here is precisely \textit{not} intended to rely on precedent, but instead to motivate a change in existing laws that govern the incipient agentic web. We also are not placing all our hopes in the kinds of judicial or administrative action that \textit{Trinko} renders unlikely. What judicial action we advocate is aimed narrowly at compelling platforms to be more explicit about how they handle agents. We hope that this candor will enable, not hinder, the competition that serves users and the broader economy. 

Platforms' and publishers' legitimate interests in their infrastructure's stability, fraud prevention, and selective contracting are intended to be defended by the proportionality principle (\S\ref{sec:proportional_restrictions}). If the property-rights argument is based not on issues defended by the proportional restriction principle, but instead simply on platforms and publishers' claimed right to determine the means of information access, we note that blocking user-authorized agents sits uneasily with the legality of screen readers and ad blockers --- both of which are forms of user-side intermediation that the law has long protected. In addition, platforms generally tend to present themselves as open and neutral fora for two-sided market participation --- this allows them, for example, to resist liability for dangerous user behavior on the platform \citep{gillespie2017platforms}. But that kind of neutrality comes at a cost: they shouldn't be able to determine the means by which users access the content that they host. 

In March of 2026, TomWikiAssist --- an AI agent built on Anthropic's Claude --- started editing Wikipedia pages and generating new ones from scratch. Human, volunteer reviewers quickly flagged the bot, it identified itself as an AI, and was banned from the website \citep{the_wikipedian_tomwikiassist_2026}. This incident exemplifies a type of concern that might be raised given the prospect of an AI agent driven future. While the Tom bot was relatively well-behaved, self-identified, and stopped (eventually), the speed and scale at which AI agents might act on the internet stands to overwhelm the structures that govern interactions on the web, many of which were designed with humans in mind. One Tom bot may be no serious issue, hundreds of them might quickly overwhelm the volunteers who review Wiki pages. In its extreme form, this objection may go so far as to revive the ``dead internet'' theory as a threat of what an agent rich internet might become --- i.e. a place where humans no longer spend time and agents alone do all of the interaction and production that we humans once did. 
\subsubsection{AI Agents Will Overwhelm Platforms Built for Human Users.}

Again though, proportional restrictions allow for blocking where harm is demonstrated. The force of this kind of objection comes from the obvious harm induced in it. If this harm is clear and present, then proportional restrictions apply. We are arguing against indiscriminate exclusion, not for a right to participate however the agent likes. Wikipedia's reaction to the Tom bot is fully compatible with our principles. The dead internet theory is a different issue that we have also countenanced in our proportional restrictions. The internet may readily handle an influx of traffic from AI agents such that no proportional restrictions based on cost or volume are incurred. And yet such an internet may move at such a pace that human use of it becomes increasingly impossible in spaces where we have strong interests in keeping AI agents out. Our proportional restriction allows that there may be spaces online whose core interests are in conflict with an influx of AI agents. Restrictions can be legitimately invoked to keep agents out of these spaces, and our transparency requirements take steps to ensure that entry under spoofed credentials is illicit entry. 

\subsubsection{Access to Online Properties Is a Small Part of the Picture.}

Another line of objection might argue that we have not gone far enough. The delegation principle does not concern only access to online properties; if applied consistently, it would also mean that AI agents should have access to any data that their principal is entitled to access. This would include, for example, user data stored by SaaS companies. Indeed, many such companies have been taking action, over the last few months, to block agent access in just the same way as was first done by online platforms \citep{mclaughlin_sap_2026}. Debates over agent access to the internet might ultimately be little more than a sideshow in comparison with these much larger markets. 

In our view, that broader conversation needs to take place too. We begin with agent access to online properties because it is salient and consequential. But we think the same principles apply to other kinds of digital access --- at the very least, we need to ensure there is a competitive market for digital infrastructure of all kinds, and that any moves to block agents can be met with customer exit to work with another firm that does not apply the same restrictions. 

\subsection{Concerns about AI Agents}
\label{sec:agent_based_concerns}

\subsubsection{Autonomous Agents Are Dangerous.}

\citet{mitchell_fully_2025} and many others argue that fully autonomous AI agents should not be developed. Broadly speaking, an ecosystem of user-authorized agents could create accountability problems, expand the attack surface for fraud and prompt injection, and increase multi-agent risks such as collusion or cascading failure \citep{dafoe_open_2020,hammond_multi-agent_2025,gabriel_ethics_2024}. 

These concerns are genuine but it is important to note that what we are advocating for here is not an unbridled, autonomous agentic-web. Instead, the agent access we advocate presupposes operation within delegated scope, with authenticated credentials, under fiduciary obligations to their principals \citep{chan_ids_2024,south_authenticated_2025,benthall_designing_2023,whitt_reweaving_2024}. Clear and present dangers --- not limited to ongoing alignment challenges, multi-agent collusion, cascading failures, emergent network effects \citep{hammond_multi-agent_2025, dafoe_open_2020} --- remain for agentic AI systems. We do not pretend that the solution to these problems will be easy or that it is inevitable with near-term technological improvements. But we have seen remarkable progress in these areas and see massive investment in the kinds of agents that could power such a future: the question is not quite ``should we build agentic AIs'' then, but ``should we restrict access to platform walled gardens or allow for user-centered universal interoperability'' \citep{kapoor_build_2025,marro_llm_2025}.

\subsubsection{Delegated AI Agent Transactions Will Introduce New Sites of Inequality.}
There is some early evidence that agent negotiations between AI agents of varying power and capabilities result in predictable losses for the ill-equipped agent \citep{sharp2025agentic, anthropic_project_2026,imas_agentic_2025,zhu_automated_2025}. This dynamic, if it persists, risks undermining the promise of AI agent advocacy and stands to fuel social inequality insofar as those with resources might, through their agents, come to be able to dominate less-resourced agents and their principals. Advocates of AI agent mediated transactions recognize this problem and suggest remedies such as baseline services \citep{krier_coasean_2025}, and agent credentialing to ensure that one's own agents interact only with others of similar capability \citep{kapoor_build_2025, chan_infrastructure_2025} but the solution to this problem is not likely to come easily. 

Agentic inequality is indeed a concern; whether it is more troubling than the existing modalities of inequality is an open question. Either way, we note that while AI agents might facilitate new modalities of inequality, they have clear potential to profoundly mitigate others. In particular, online platforms currently have tremendous power over their users \citep{lazar_governing_2025}. AI agents won't solve every social ill. But they could do a great deal to solve that one. 

\section{Conclusion}
\label{sec:conclusion}

AI agents are now capable enough to act on behalf of users across much of their online lives. And yet the normative infrastructure required to support this action is conspicuously lacking: currently, agents are governed and constrained by a patchwork built on the pattern of earlier technologies. Autonomous AI systems that could significantly enhance their principals' well-being are subject to legacy anti-bot regulations and practices, calibrated for a different technology. This paper has argued, first, that we urgently need to hold an open, societal conversation about the proper scope of AI agent access to the web (as well as to other digital surfaces). Second, we defended three normative principles that should be central in this conversation: (1) a delegation principle in which users ought to ordinarily be entitled to access services through user-authorized agents, (2) two-way transparency in which these agents must disclose their principals and purpose, while platforms must disclose how they handle AI agents, and (3) proportional restriction in which platforms must justify restrictions with reference to concrete or reasonably expected harms. We then laid out legal pathways that might support these principles. Third, we identified some minimal paths by which administrative decision and legislative action might better facilitate an open, decentralized economy of agent-advocates. 

\newpage


\bibliographystyle{plainnat}
\bibliography{agentic_web}
\end{document}